\documentclass[preprint,
preprint,
 amsmath,amssymb,
 aps, physrev,
pra,
]{revtex4-2}

\usepackage{graphicx}
\usepackage{dcolumn}
\usepackage{bm}

\usepackage{float}
\begin{document}

\preprint{}

\title{\textbf{Temperature-driven enhancement and sign reversal of field-like torque in Py/FePS$_3$ bilayers} 
}%

\author{Dhananjaya Mahapatra}
\email{Contact author: dm20rs019@iiserkol.ac.in}
 \author{Anudeepa Ghosh}
\author{ Harekrishna Bhunia}
  \author{Bipul Pal}
  \author{Partha Mitra}
 
\affiliation{%
 Department of Physical Sciences,
 Indian Institute of Science Education and Research Kolkata, Mohanpur, West Bengal,741246, INDIA.
}
\begin{abstract}
Electrical manipulation of magnetization via current-induced spin–orbit torques (SOTs) offers a promising route toward nonvolatile and energy-efficient spintronic devices. In this work, we present a comprehensive investigation of SOTs in Py/FePS$_3$ bilayer devices, where Py/FePS$_3$ is a layered van der Waals antiferromagnetic insulator. Using low-frequency harmonic Hall measurements, we quantify both field-like and damping-like torque components and examine their dependence on temperature. We find that interfacing Py with Py/FePS$_3$ leads to a pronounced enhancement of the field-like torque efficiency compared to Py reference devices, while the damping-like torque remains largely unaffected. Strikingly, the field-like torque efficiency exhibits a strong temperature dependence, including a clear sign reversal upon cooling. This behavior occurs despite negligible charge-current flow through the Py/FePS$_3$ layer, indicating that the observed torque modulation arises from interfacial effects rather than bulk transport. The close correlation between the temperature evolution of the field-like torque and the antiferromagnetic ordering of Py/FePS$_3$ highlights the active role of antiferromagnetic insulators in controlling spin–orbit torque symmetry and efficiency, and suggests new pathways for torque engineering in magnetic heterostructures.

\end{abstract}

\maketitle

\section{\label{sec:level1}INTRODUCTION:}
Electrical control of magnetization is a key requirement for the realization of next-generation nonvolatile and energy-efficient spintronic memory devices. Among the various approaches, current-induced spin–orbit torque (SOT) has emerged as a particularly powerful mechanism, in which an in-plane charge current flowing through a material with strong spin–orbit interaction exerts a torque on an adjacent magnetic layer \cite{PhysRevB.90.224427, shao2021roadmap}. Such torques can originate from bulk spin–orbit effects, most notably the spin Hall effect \cite{PhysRevLett.83.1834, PhysRevLett.85.393, PhysRevLett.92.126603,6516040}, where charge carriers with opposite spin angular momentum are transversely deflected in opposite directions, leading to a spin current perpendicular to the charge flow. In addition to bulk contributions, interfacial spin–orbit coupling can also play a significant role; for example, the Rashba–Edelstein effect generates a non-equilibrium spin accumulation at structurally asymmetric interfaces \cite{nakayama2016rashba, edelstein1990spin}, which can exert a torque on the neighboring magnetic moments \cite{mihai2010current}. Recent studies have further demonstrated that SOTs can be generated directly within metallic ferromagnets and antiferromagnetic materials, highlighting the importance of magnetic order in governing both the magnitude and symmetry of the resulting torques \cite{schippers2020large, shi2023recent}. In particular, the presence of antiferromagnetic order introduces additional pathways for spin–angular-momentum transfer, offering new opportunities for tuning SOT efficiency through magnetic correlations and interfacial exchange interactions \cite{schippers2020large}.
The emergence of layered van der Waals (vdW) materials has opened new pathways for engineering interfacial spin–orbit phenomena in magnetic heterostructures \cite{tang2021spin, shi2023recent, zhao2025novel, ryu2024van}. Among these materials, vdW antiferromagnets (AFMs) have attracted significant attention because they combine the intrinsic advantages of antiferromagnetism—absence of stray fields, ultrafast spin dynamics, and robustness against external magnetic perturbations—with the structural benefits of vdW systems such as atomically sharp interfaces and minimal interlayer hybridisation \cite{shao2021roadmap}. These properties provide an ideal platform for studying interfacial spin transport, proximity coupling, and torque generation at the nanoscale. A large class of vdW antiferromagnets is represented by the MPX$_3$ family (M = Mn, Fe, Ni, Co; X = S, Se), where FePS$_3$ is a prototypical type-C antiferromagnet characterized by strong in-plane magnetic anisotropy and a layered honeycomb-like structure \cite{matsuoka2024mpx, coak2019isostructural, masubuchi2008phase}. The review \cite{shi2023recent} highlights that layered AFMs such as FePS$_3$ and MnPS$_3$ support tunable magnon excitations, long-distance magnon transport, and robust antiferromagnetic ordering even in ultrathin form. Their insulating nature eliminates shunting of charge current and enables a clean environment to explore pure spin- and orbital-mediated interfacial torques. Moreover, recent studies show that vdW AFM/FM heterostructures can significantly modify the magnetic properties of adjacent ferromagnets through proximity coupling. For example, FePS$_3$ interfaced with a ferromagnet such as Fe$_3$GeTe$_2$ has been shown to enhance coercivity and increase the Curie temperature \cite{dai2021enhancement}, demonstrating strong interlayer exchange interactions across vdW gaps. These proximity-driven modifications indicate that vdW AFMs can actively tune the interfacial spin configuration and thereby influence the efficiency and symmetry of spin–orbit torques (SOTs). The review also reports that AFM-based heterostructures can generate appreciable damping-like and field-like torque components, as observed in systems such as NiPS$_3$/Py \cite{schippers2020large}, where large spin Hall conductivity and efficient spin–orbit torque generation have been measured. This suggests that vdW AFMs are not passive layers; instead, they actively participate in spin conversion processes through interfacial spin mixing, orbital hybridization, and magnon-mediated transfer.

In this work, we investigate current-induced spin–orbit torques in FePS$_3$/Py bilayer devices using harmonic Hall measurements. Our results reveal enhancement of the interfacial field-like torque in the presence of the layered antiferromagnetic insulator FePS$_3$. Systematic temperature-dependent measurements show a strong evolution of the field-like torque with decreasing temperature, including a sign reversal at low temperatures. In contrast, the damping-like torque remains essentially unchanged over the same temperature range. The distinct temperature dependence and sign change of the field-like torque strongly indicate that the observed behavior originates from the magnetic ordering of FePS$_3$, highlighting the crucial role of antiferromagnetic correlations in modulating interfacial spin–orbit torque symmetry.

\section{Device fabrication and Experimental method}

\begin{figure*}[h!]
\includegraphics[width=1\textwidth]{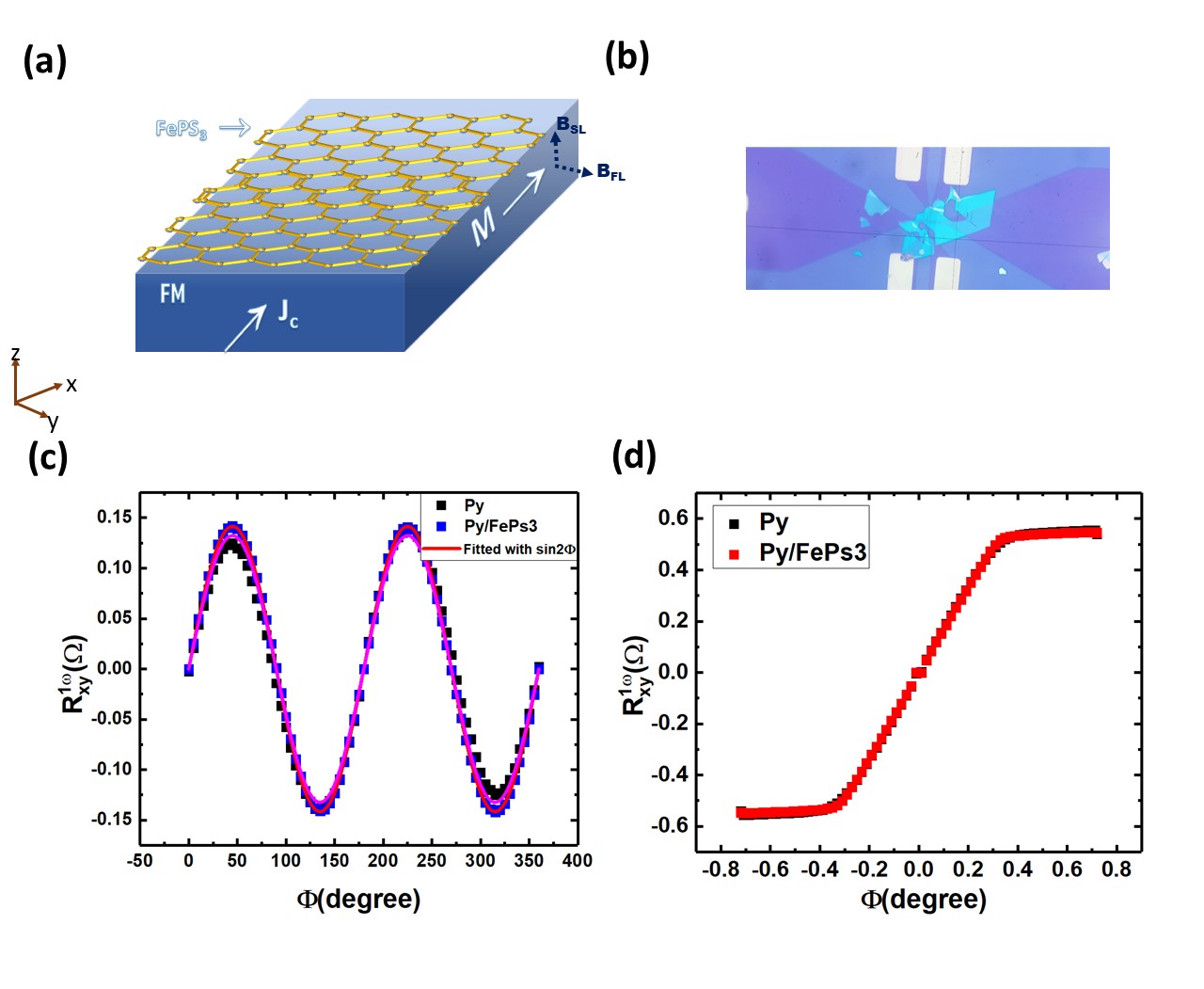}
    \caption{(a) Schematic illustration of the FePS$_3$/Py bilayer device geometry and measurement configuration. An in-plane charge current $J_c$flowing through the Py layer generates spin–orbit torques, giving rise to field-like ($B_{FL}$) and damping-like ($B_{DL}$) effective fields acting on the magnetization M.
(b) Optical microscope image of a representative FePS$_3$/Py Hall bar device used for harmonic Hall measurements.
(c) Angular dependence of the second-harmonic Hall resistance $R_{xy}^{1\omega}$measured as a function of the in-plane magnetic field angle $\phi$for Py and Py/FePS$_3$ devices. The solid lines represent fits using a $sin(2\phi)$ dependence.
(d) First-harmonic Hall resistance 
$R_{xy}^{1\omega}$ as a function of 
$\phi$, showing comparable magnetic response for Py and Py/FePS$_3$, indicating that the static magnetic properties of Py remain largely unaffected by the FePS$_3$ layer.}
    \label{1}
\end{figure*}
For this study, ferromagnetic (FM) Hall-bar devices were fabricated on Si/SiO$_2$ (300 nm) substrates using electron-beam lithography. The device geometry consisted of standard Hall-bar structures with current-channel dimensions of 2 × 30 $\mu m^2$ and voltage pickup lines of 0.5 $\mu m$, patterned using the ELPHY Quantum system integrated with the Raith electron-beam lithography platform. Following lithography, a 10 nm-thick Py (NiFe) ferromagnetic layer was deposited by thermal evaporation at a deposition rate of 0.8–0.9 $A^0/s$, ensuring uniform film growth and smooth interfaces.

The antiferromagnetic vdW material FePS$_3$ was prepared by mechanical exfoliation, in which bulk FePS$_3$ crystals were repeatedly cleaved using standard Scotch tape until thin, optically homogeneous flakes were obtained. These exfoliated flakes were then transferred onto the pre-patterned NiFe Hall-bar devices using a micromanipulator-based transfer stage. During the transfer process, the substrate was maintained at $100^0$ C to promote good adhesion of the FePS$_3$ flake and to minimise trapped contaminants at the vdW/FM interface. This deterministic transfer procedure enabled precise placement of FePS$_3$ directly on top of the Py channel, ensuring clean, atomically sharp vdW interfaces essential for probing interfacial field-like torque enhancement. To quantify the current-induced torques generated in the FePS$_3$/NiFe bilayers, we employed the second-harmonic Hall measurement technique, which is a sensitive and widely used method for separating field-like and damping-like torque contributions in ferromagnetic systems. A low-frequency sinusoidal current, $I(t)= I_0 sin(\omega t)$.
was applied along the longitudinal channel of the Hall-bar device. The resulting voltage signals were detected simultaneously at the fundamental frequency ($R^{1\omega}$) and at the second-harmonic frequency ($R^{2\omega}$) using a lock-in amplifier. The first-harmonic resistance primarily reflects the magnetoresistive response of the device, whereas the second-harmonic voltage contains information about current-induced torques, Oersted field contributions, and possible thermoelectric effects arising from Joule heating. During measurement, an external magnetic field was applied in-plane and rotated through $360^0$, allowing the angular dependence of both $R^{2\omega}_{xy}$ (transverse) signals to be recorded. The transverse second-harmonic signal exhibits characteristic $cos\phi$ and $cos2\phi cos\phi$ symmetries that correspond to the field-like torque and damping-like torque/Oersted components, respectively. By fitting these angular dependencies with the standard harmonic Hall model, the effective field associated with the field-like torque $B_{FL}$ was extracted. The longitudinal second-harmonic voltage was simultaneously analysed to separate thermoelectric artifacts such as the anomalous Nernst effect (ANE) or spin Seebeck effect (SSE), ensuring that only genuine torque contributions were incorporated into the final analysis. Measurements were performed over a wide temperature range using a closed-cycle cryostat, enabling us to probe how the interfacial torque evolves as the antiferromagnetic order in FePS$_3$ becomes stronger at low temperatures. The harmonic technique therefore provides a direct and quantitative link between interfacial magnetic coupling in FePS$_3$/NiFe bilayers and the observed enhancement of field-like torque.

\section{Results and Discussion}

\begin{figure}[h!]
\includegraphics[width=1\textwidth]{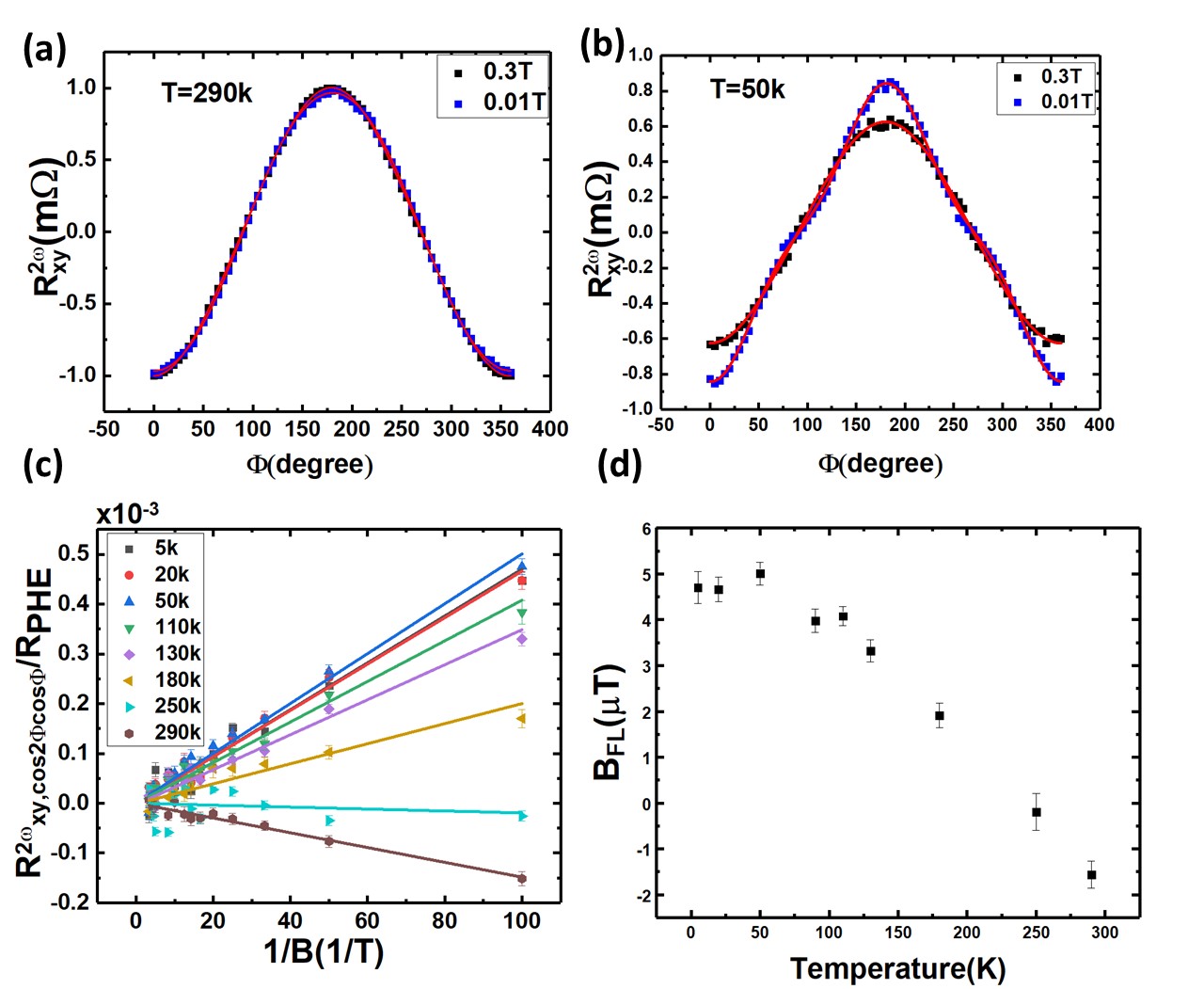}
    \caption{(a,b) Angular dependence of the second-harmonic Hall resistance $R_{xy}^{2\omega}$
 measured at different in-plane magnetic fields (0.3 T and 0.01 T) at (a) T=290 K and (b) T=50 K for the FePS$_3$/Py bilayer device. The solid lines represent fits based on the standard harmonic Hall model. (c) Extracted quantity 
$R^{2\omega}_{xy,cos2\phi cos\phi}/R_{PHE}$ plotted as a function of inverse magnetic field 
1/B at various temperatures, where the linear dependence is used to separate the field-like torque contribution.
(d) Temperature dependence of the effective field-like effective field ($B_{FL}$) , showing a pronounced evolution and sign reversal upon decreasing temperature, while the damping-like torque remains negligible within experimental uncertainty shown in fig. \ref{4}.}
    \label{2}
\end{figure}

Harmonic Hall measurements were performed using a low-frequency excitation of 13 Hz with a lock-in amplifier by applying an oscillating current of 5 mA to the device. Both the first-harmonic (1$\omega$) and second-harmonic (2$\omega$) transverse Hall resistances were simultaneously detected. During the measurements, by using vector magnet an external magnetic field was rotated in the plane of the sample from $0^0$ to $360^0$ with respect to the current channel, while the field magnitude was varied between 0.3 T and 0.01 T. The first-harmonic transverse Hall resistance$R_{xy}^{1\omega}$ represents a linear response that depends solely on the equilibrium orientation of the magnetization. The in-plane component of the magnetization gives rise to the planar Hall effect (PHE), while the out-of-plane component contributes through the anomalous Hall effect (AHE). As a result, $R_{xy}^{1\omega}$ depends on the relative angle between the magnetization and the applied current, which is controlled by the external magnetic field $B_{ext}$. In contrast, the second-harmonic Hall resistance $R_{xy}^{2\omega}$ originates from the oscillatory motion of the magnetization about its equilibrium direction, driven by current-induced spin–orbit torques. These torques arise from spin currents injected into the ferromagnetic layer from the adjacent layer, which can be generated either through bulk mechanisms such as the spin Hall effect or through interfacial effects such as the Rashba–Edelstein effect. Because the second-harmonic voltage is nonlinear in current, it is customary to extract effective magnetic fields normalized by the applied current. These effective fields quantify the strength of the field-like and damping-like spin–orbit torques per unit magnetization and are commonly referred to as torque efficiencies.

Figures 1(c) and 1(d) show the first-harmonic Hall response of the devices. In Fig. 1(c), the transverse resistance measured under rotation of an in-plane magnetic field exhibits a clear angular dependence that is well fitted by the established planar Hall effect (PHE) expression \cite{mcguire2003anisotropic},
\begin{equation}
    R^{1\omega}_{xy}= R_{PHE} sin(2\phi)
\end{equation}

The planar Hall effect originates from spin-dependent scattering in the ferromagnetic layer and reflects the in-plane orientation of the magnetization relative to the current direction. Figure 1(d) shows the first-harmonic Hall resistance measured as a function of an out-of-plane magnetic field. The data are well described by linear fits in the high-field saturation regime, and the intersection point of these linear branches yields the anomalous Hall resistance $R_{AHE}$
\begin{figure}[h!]
\includegraphics[width=0.8\textwidth]{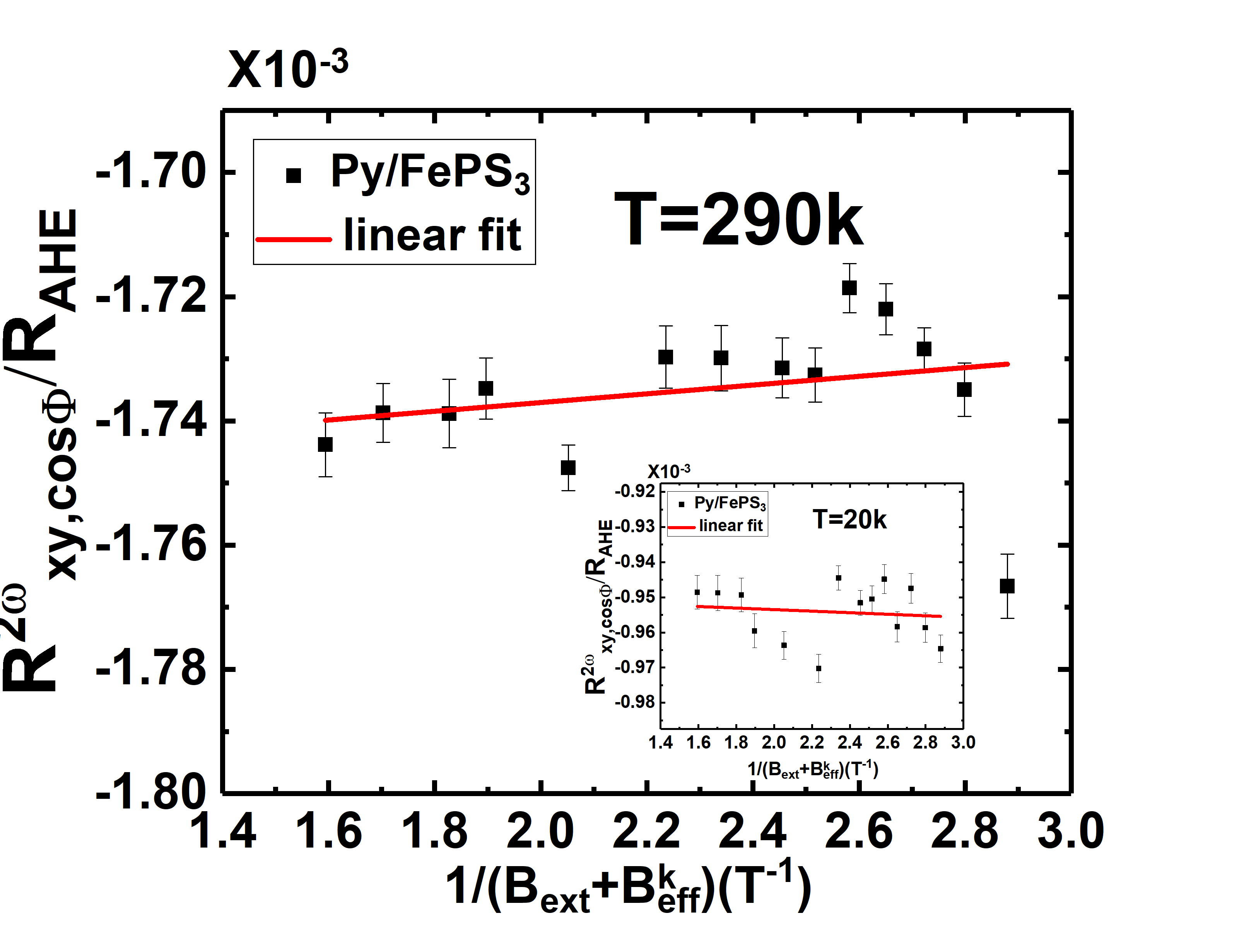}
    \caption{Normalized second-harmonic transverse resistance $\frac{R^{2\omega}_{xy,cos\phi}}{R_{AHE}}$ plotted as a function of 
$\frac{1}{B_{ext}+B^k_{eff}}$ for the Py/FePS$_3$ bilayer at 
T=290 K. Symbols represent the experimental data and the solid line is a linear fit. The nearly vanishing slope indicates a negligible SL like effective field contribution within the experimental uncertainty.
Inset: Corresponding measurement at
T=20 K. The absence of any appreciable linear dependence confirms that the SL like torque contribution remains negligible over the investigated temperature range.}
    \label{4}
\end{figure}
. A direct comparison of both the planar Hall effect (Fig. 1(c)) and anomalous Hall effect (Fig. 1(d)) signals for the Py single-layer and Py/FePS$_3$ bilayer devices reveals nearly identical responses. This close overlap indicates that the contribution of charge-current flow through the FePS$_3$ layer is negligible, consistent with its insulating nature, and confirms that the measured Hall signals predominantly originate from the Py layer.

We next analyze the second-harmonic Hall resistance $R_{xy}^{2\omega}$, by rotating the external magnetic field in the plane of the sample with respect to the current channel. The angular dependence of $R_{xy}^{2\omega}$ is fitted using the previously established harmonic Hall model which allows the separation of the field-like and damping-like torque contributions and the expression is give by   \cite{aoki2023gigantic, mahapatra2025evidence}
\begin{equation}
\begin{split}
    R_{xy}^{2\omega}(\Phi)=[R_{AHE}(\frac{B_{SL}}{B_{ext}+B_{eff}^k})+\alpha B_{ext} + R_{\nabla T}]cos(\Phi)\\
    +2R_{PHE}(\frac{B_{FL}+B_{Oe}}{B_{ext}})cos(2\Phi)cos(\Phi)
    \end{split}
    \label{eq1}
\end{equation}
\begin{figure}[h!]
\includegraphics[width=1.1\textwidth]{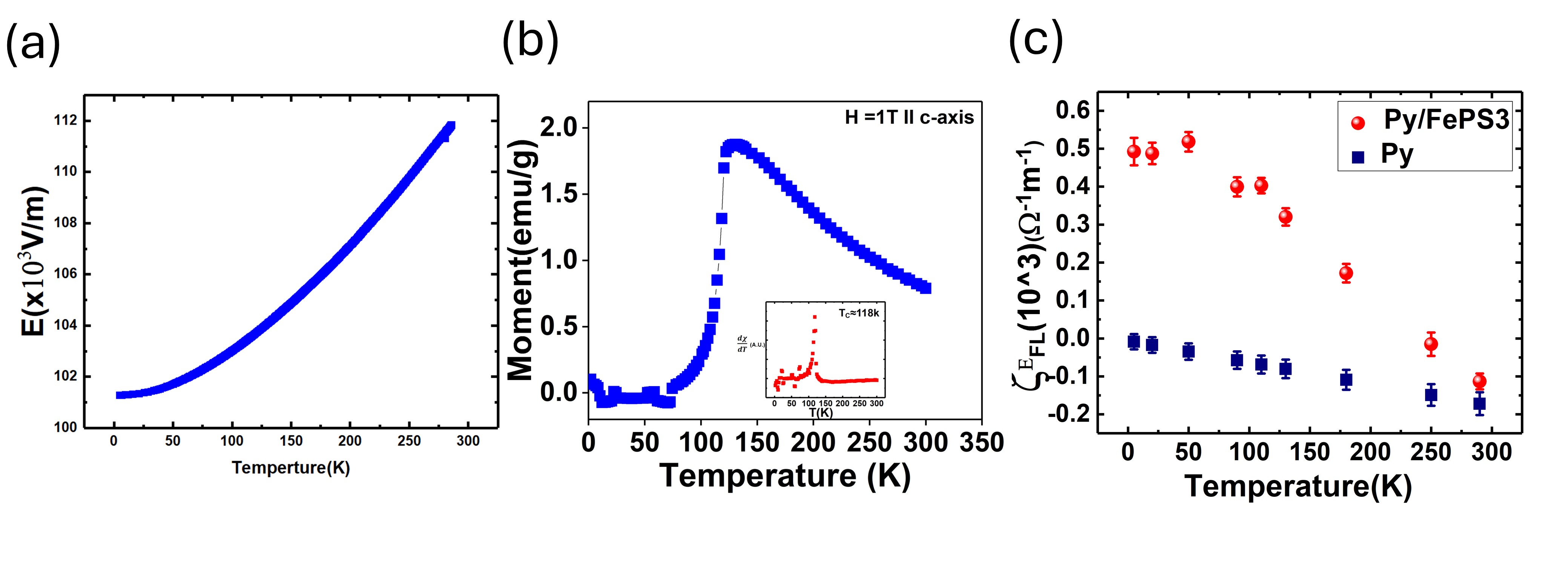}
    \caption{(a) Temperature dependence of the electric field across the device calculated for an applied current of 5 mA, showing a monotonic decrease with decreasing temperature, consistent with metallic transport dominated by the Py layer. Temperature-dependent magnetization of FePS$_3$ measured using a SQUID magnetometer under an applied magnetic field of 1 T, with the anomaly near T=118 identifying the antiferromagnetic Néel temperature of FePS$_3$.(c) Temperature dependence of the field-like torque efficiency $\zeta_{FL}^E$ for Py/FePS$_3$ (red circles) and Py reference devices (blue squares). The pronounced enhancement and sign reversal of $\zeta_{FL}^E$ in Py/FePS$_3$ correlates strongly with the antiferromagnetic ordering of FePS$_3$, while the Py-only device shows a weak and monotonic temperature dependence.}
    \label{3}
\end{figure}
Figure 2(a) shows the second-harmonic Hall response measured at room temperature under different in-plane magnetic fields. The angular modulation of $R_{xy}^{2\omega}$ 
is relatively weak, indicating modest field-like and damping-like torque components. In contrast, upon lowering the temperature, the amplitude of the second-harmonic signal increases markedly, as shown in Fig. 2(b). The substantially enhanced angular variation at low temperature highlights a strong modification of the spin–orbit torque response compared to room temperature. The damping-like (spin–orbit) torque contribution was extracted by plotting the coefficient of the $cos\phi$ term as a function of $\frac{1}{B_{ext}+B^k_{eff}}$. Linear fitting of this dependence yields a negligible slope, indicating that the damping-like torque contribution is minimal within experimental uncertainty. In contrast, the field-like torque contribution was obtained by plotting the coefficient of the $cos2\phi cos\phi$ term as a function of $1/B_{ext}$. From the linear fits, a clear difference in slope is observed between the Py single-layer and the $Py/FePS_3 $ bilayer devices, as shown in the Supplementary Information. This difference reflects an additional interfacial contribution to the field-like effective field arising from the presence of the FePS$_3$ layer. We further performed a systematic temperature-dependent analysis of the field-like torque. As shown in Fig. 2(c), the slope of the $cos2\phi cos\phi$ coefficient plotted against $1/B_{ext}$ exhibits a strong temperature dependence, where the slope directly corresponds to the field-like effective field. The extracted field-like effective field as a function of temperature is summarized in Fig. 2(d). Two key features are evident: first, a pronounced enhancement of the field-like effective field upon decreasing temperature; and second, a clear sign reversal from negative to positive values. Notably, this sign change emerges upon introducing the FePS$_3$ layer, despite the negligible charge-current flow through FePS$_3$, indicating that the observed behavior originates from interfacial effects rather than bulk charge transport. The magnitude of the extracted effective fields generally depends on experimental parameters such as the applied electric field, the thickness of the ferromagnetic layer, and the saturation magnetization. To enable a meaningful comparison between different devices and measurement conditions, we therefore evaluate the field-like torque efficiency, which provides a normalized measure of the SL or FL effective field. The SL or FL torque efficiency is defined as \cite{PhysRevB.92.064426}
\begin{equation}
    \zeta_{SL/FL}^E=\frac{2e}{\hbar} M_s t_{\small{FM}} \frac{B_{SL}}{E}
\end{equation}
Figure 3(a) shows the temperature dependence of the electric field in the device, calculated for an applied current of 5 mA. As the temperature is reduced, the device resistance decreases monotonically, leading to a corresponding reduction in the electric field across the entire temperature range. This temperature-dependent electric field is taken into account in the normalization procedure for the extracted torque efficiencies. Specifically, the electric field at each temperature is used to normalize the corresponding field-like effective field values. The ferromagnetic layer thickness is approximately 10 nm, and the saturation magnetization $M_s$ is determined from the saturation value of the anomalous Hall effect (AHE) measurements. Using these parameters, we calculate the field-like torque efficiency for both the Py single-layer and the Py/FePS$_3$ bilayer devices, as summarized in Fig. 3(c). Compared to the Py reference, the Py/FePS$_3$ bilayer exhibits a significantly enhanced field-like torque efficiency over the entire temperature range, with a pronounced increase and sign reversal at lower temperatures.

 In a recent study, Schippers et al. \cite{schippers2020large} reported finite values of both damping-like and field-like torque efficiencies in NiPS$_3$/ferromagnet bilayers, which were attributed to interfacial magnetization alignment and spin–orbit coupling at the vdW AFM/FM interface. In contrast, in the present FePS$_3$/Py bilayer system, we observe a qualitatively different behavior: the damping-like torque contribution remains negligible within experimental uncertainty, while a finite and strongly temperature-dependent field-like torque is clearly resolved. The presence of a sizable field-like torque, despite the absence of a measurable damping-like component, indicates that the dominant torque generation mechanism in FePS$_3$/Py originates from interfacial spin–orbit effects rather than bulk spin-current injection. This observation is consistent with theoretical predictions suggesting that antiferromagnetic insulators can generate predominantly field-like torques through interfacial exchange coupling, orbital hybridization, and spin-dependent scattering, without requiring significant charge-current flow or spin Hall–driven damping-like contributions \cite{amin2016spin,amin2016spin1}. 

 To elucidate the origin of the observed field-like torque, we first examine the charge-current distribution in the FePS$_3$/Py bilayer. As shown in Figs. 1(c) and 1(d), both the planar Hall effect (PHE) and anomalous Hall effect (AHE) responses of the FePS$_3$/Py bilayer closely overlap with those of the Py single-layer device. This observation indicates that the static magnetotransport properties of Py remain unchanged upon interfacing with FePS$_3$ and, importantly, confirms that charge-current shunting through the FePS$_3$ layer is negligible due to its highly insulating nature. Given the much higher resistivity of FePS$_3$ compared to metallic Py, the applied current predominantly flows through the Py layer. A possible contribution to the field-like torque could arise intrinsically from the Py layer itself. To address this, we performed identical harmonic Hall measurements on Py single-layer devices. As shown in Fig. 3(c), the extracted field-like torque efficiency of Py alone is significantly smaller than that of the FePS$_3$/Py bilayer. Moreover, the temperature dependence of the field-like torque in the Py single-layer device is weak, exhibiting negligible variation between room temperature and low temperature. In stark contrast, the FePS$_3$/Py bilayer shows an approximately fivefold enhancement of the field-like torque at low temperature. The pronounced enhancement and strong temperature dependence observed exclusively in the FePS$_3$/Py bilayer demonstrate that the field-like torque does not originate from the Py layer alone. Instead, it arises from interfacial effects induced by the presence of the FePS$_3$ layer. These results strongly support an interfacial spin–orbit torque mechanism associated with the antiferromagnetic FePS$_3$/Py interface, rather than bulk charge-current or intrinsic ferromagnetic contributions.

To further explore the role of the antiferromagnetic phase transition in FePS$_3$, we performed temperature-dependent magnetization measurements using a SQUID magnetometer. As shown in Fig. 3(b), the magnetization data clearly identify the Néel temperature of FePS$_3$ at approximately 118 K. The temperature dependence of the electric field [Fig. 3(a)] shows a monotonic decrease upon cooling, reflecting the metallic transport behavior of the bilayer device. Given the insulating nature of FePS$_3$, this behavior indicates that the overall resistance of the bilayer is dominated by the metallic Py layer. Consistent with this interpretation, the saturation magnetization $M_s$
 extracted from anomalous Hall effect (AHE) measurements exhibits no anomaly near 118 K, further confirming that the static magnetic and transport properties are governed primarily by the Py layer. In contrast, the field-like torque in the Py/FePS$_3$ bilayer displays a markedly different temperature dependence. As shown in Fig. 3(c), the field-like torque efficiency undergoes a sign reversal at around 250 K, followed by a pronounced increase in magnitude upon further cooling toward the Néel temperature of FePS$_3$. Below approximately 118 K, the field-like torque shows a relatively weak temperature dependence down to about 90 K, before increasing again at lower temperatures down to 5 K. This non-monotonic temperature evolution, which is absent in Py single-layer devices, strongly suggests that the field-like torque is governed by interfacial effects associated with the magnetic ordering of FePS$_3$ rather than by bulk transport or intrinsic ferromagnetic properties of Py.

\section{Conclusion}

In conclusion, we have investigated current-induced spin–orbit torques in Py/FePS$_3$ bilayer devices using harmonic Hall measurements, with a particular focus on the role of a layered antiferromagnetic insulator. We demonstrate that interfacing Py with FePS$_3$ leads to an enhancement of the field-like torque efficiency, while the damping-like torque remains negligible within experimental uncertainty. Systematic temperature-dependent measurements reveal a strong evolution of the field-like torque, including a clear sign reversal and a non-monotonic temperature dependence that correlates with the antiferromagnetic ordering of FePS$_3$. Importantly, transport and magnetization measurements confirm that charge-current flow and static magnetic properties are dominated by the metallic Py layer, indicating that the observed torque modulation originates from interfacial effects rather than bulk transport in FePS$_3$. The absence of a comparable temperature dependence in Py single-layer devices further supports this conclusion. Our results highlight the active role of antiferromagnetic insulators in selectively controlling field-like spin–orbit torques and provide a viable route for engineering torque symmetry and efficiency through interfacial magnetic order in van der Waals heterostructures.

\begin{acknowledgments}
The authors thank IISER Kolkata, an autonomous research and teaching institution funded
by the MoE, Government of India, for providing the financial support and infrastructure. The authors also thank CSIR and UGC for providing fellowship. AG acknowledges IISER Kolkata and ANRF (PDF/2025/005009) for fellowships.

\end{acknowledgments}
\section*{DATA  AVAILABILITY}
The data that support the findings of this article are not
publicly available. The data are available from the authors
upon reasonable request.
\appendix

\nocite{*}
\newpage
\bibliography{apssamp}

\end{document}